# Relaxation dynamics in reverse Monte Carlo

Akash Kumar Ball[#], Suhail Haque[#] and Abhijit Chatterjee*

Department of Chemical Engineering, Indian Institute of Technology Bombay, Mumbai 400076, India

*Email: abhijit@che.iitb.ac.in

[#] Equal contribution

**Abstract**

The reverse Monte Carlo (RMC) method is widely used in structural modelling and analysis of experimental data. More recently, RMC has been applied to the calculation of equilibrium thermodynamic properties and dynamic problems. These studies point to the importance of properly converging RMC calculations and understanding the relaxation behavior in RMC. From our detailed RMC calculations, we show that the relaxation comprises of both fast and slow aspects. A metric is introduced to assess whether fast equilibration is achieved, i.e., detailed balance condition is satisfied. The metric, essentially an equilibrium constant for RMC, is used as a test for quasi-equilibration. The slow evolution is analogous to glassy materials, i.e., it is characterized empirically in terms of the Kohlrausch-Williams-Watts (KWW) function, i.e., stretched exponentials. This feature can be exploited to estimate the convergence error or to extrapolate statistical quantities from short RMC calculations.

1. Introduction

The Reverse Monte Carlo (RMC) method[1] is a variation of the Metropolis-Hastings algorithm. It has been widely employed for structural modelling of aqueous solutions[2,3], simple liquids[4], molecular liquids[5], molten salts[6], metallic[7] and covalent[8] glasses, crystalline materials[9] and microporous carbon[10,11], investigation of diffraction patterns for polycrystalline materials[12], analysis of magnetic structures[13], estimation of the pair interaction from radial distribution function (rdf)[14,15] and analysis of EXAFS data[16]. In these studies, the goal usually is to construct material structures that are consistent with the available experimental rdf or scattering data. Recently, we have extended the use of RMC to thermodynamic[17–20] and dynamic problems[21]. In this approach, RMC does not require any experimental inputs, such as rdf or scattering data. Instead, inputs such as the short-ranged order (SRO), composition, interactions and temperature are used to determine the equilibrium/non-equilibrium atomic arrangements, configurational free energy, chemical potential and other properties of interest. Since all RMC calculations ultimately comprise of a sequence of trial moves to evolve the material structure such that target constraints are satisfied, guaranteeing convergence is a crucial part of the calculation.

A problem often reported with RMC for structural modelling applications is that there are many ways for the configuration to be arranged to fit the experimental data[22]. As it has been shown recently[19], simply fitting the data by meeting specified RMC target constraints, such as an rdf or SRO parameter, does not guarantee convergence. The material structure may continue to evolve. Several examples are provided later. However, once the RMC calculation converges the statistical distributions are uniquely obtained. One needs to therefore systematically understand the convergence behavior for RMC. A mathematical framework for this purpose is discussed in this paper.

In this study, we restrict ourselves to on-lattice structures, which we have investigated extensively in the past in terms of structure, energetics and dynamics [17–20]. The results should also be relevant to off-lattice problems. The rdf itself is a statistical feature, however, it corresponds to a moment of the probability distribution for a cluster of sites[19]. This site-cluster probability distribution continues to evolve[19], even after the simulated rdf has been matched to the target rdf. When the calculation has not converged, thermodynamic estimates based on such cluster probability distributions will depend on the length of the RMC calculation. An RMC calculation is said to be truly converged when the statistical distributions associated with the structure become stationary. A test for convergence can involve tracking changes in these distributions over a very large number of trial moves. Using this approach[19], 10-2000 million or move trial moves may be required to achieve convergence (the exact number depends also on the system size). The presence of statistical noise can make tracking changes in a distribution challenging. Similar issues exist with Metropolis Monte Carlo algorithms.

The present study was motivated by the need to develop a metric for convergence that does not require comparisons between current and previous probability distributions. The metric presented here simply assesses on-the-fly how well detailed balance is satisfied with the present configuration. The metric collapses the evolution of several site-cluster probability values into essentially a handful of equilibrium constants, which makes it a convenient mathematical tool for understanding the dynamic relaxation in RMC. The initial part of our discussion pertains to the concept of equilibrium constant in RMC. Next, we discover that analogous to MC in some cases fast relaxation is achieved and the detailed balance condition is satisfied. In addition, slow relaxation may be present, i.e., the cluster probability distributions continue to evolve while the structure is quasi-equilibrated. The relaxation dynamics can be described in terms of a stretched exponential function. This behavior can be exploited for gaining computational efficiency by extrapolating results from short RMC calculations or for estimating the error in a short RMC calculation. These aspects have not been explored previously in the RMC literature to the best of our knowledge. The usual RMC algorithm involves stopping the calculation once the target constraints are achieved.

In Section 2, we describe our RMC approach. In Section 3, the convergence criterion based on detailed balance condition is derived. In Section 4, the convergence criterion is evaluated. In Section 5, we extend our discussion to multiple short-ranged order parameters. Finally, conclusions are provided in Section 6.

## 2. Overview of the RMC approach

### 2.1 RMC algorithm

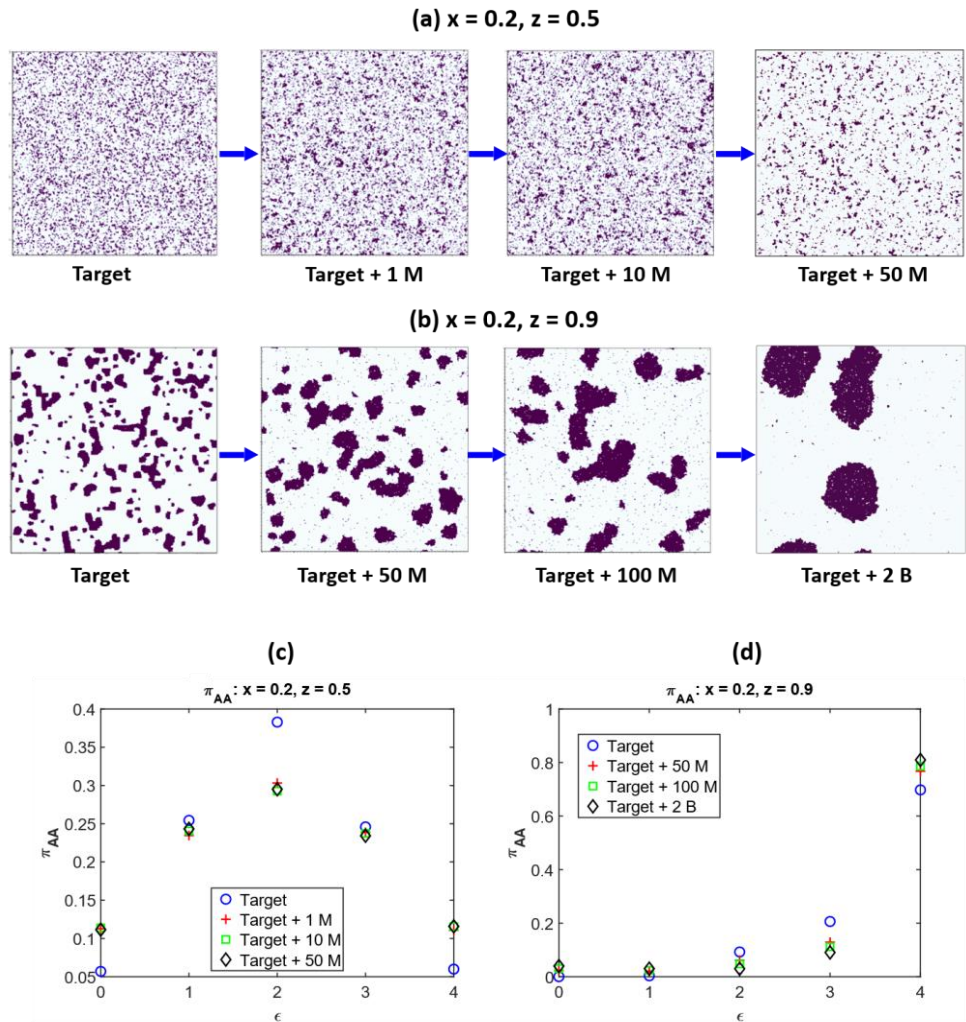

Figure 1: Relaxation in RMC: (a) $x = 0.2, z = 0.5$, (b) $x = 0.2, z = 0.9$. Variation in $\pi_{AA}(\epsilon)$ with RMC iterations: (c) $x = 0.2, z = 0.5$, (d) $x = 0.2, z = 0.9$. Results are presented for 2D square (100) lattice. 'M' and 'B' denotes million and billion trial moves, respectively.

Consider the adsorption problem involving a two-dimensional square lattice where a site is occupied by species $A$ or it is vacant. Here $V$ is used to denote vacancy. Figure 1(a-b) shows examples of RMC configurations generated. The overall coverage $x$ is specified along with the target first nearest neighbor (1NN) short-range order (SRO) parameter $z$. $x$ is the fraction of sites occupied by $A$, while $z$ is the probability of an 1NN site around an $A$-site being occupied.

The RMC algorithm used here is similar to standard implementations[1,23–25]. The initial configuration provided to RMC is a perfectly random one where the number of $A-A$ bonds is $\frac{1}{2}N_A cx$. The target number of $A-A$ bonds for the specified $z$, is $N_{AA,t} = \frac{1}{2}N_A cz$, where $N_A = xN_t$ is the number of $A$ sites and $N_t$ is the total number of sites. The RMC algorithm is as follows:

Step 1: $N_{AA}$ is calculated for current configuration. The distance from the target structure is calculated as $d_t^2 = (N_{AA} - N_{AA,t})^2$.

Step 2: Swap move: A pair of randomly selected $A$ and $V$ sites are allowed to interchange their site occupations. The new number of $A-A$ bonds in the new configuration is $N_{AA,n}$ and distance from the target structure $d_{t,n}^2$ is calculated.

Step 3: Acceptance criterion: The move is accepted with the probability:

$$p_{acc} = \min\left[1, \exp\left(-\frac{d_{t,n}^2 - d_t^2}{\theta}\right)\right]. \tag{1}$$

In the above expression, $\theta$ is a constant. The role of $\theta$ in RMC calculations is discussed in Section 2.3. If the move is accepted, new configuration remains. Otherwise, the system is returned to the previous configuration.

Step 4: Steps 1-3 are repeated till $\pi_{AA}$ and $\pi_{VA}$ are converged.

Traditionally, the RMC calculation is stopped once $d_t^2$ reaches an equilibrium value. This usually happens once the target is reached. For instance, in our calculations the value of $d_t^2$ can hover near zero once target is reached. However, from Figure 1a-b we can see that the structure is not converged at target. Some amount of coarsening is observed, with coarsening in panel b being more extensive than panel a. The fully relaxed structures from panel a and b are different although the composition $x$ is identical.

## 2.2 Short-range order (SRO) parameter and probability distributions

Figure 1a-b provides evidence of slow convergence. To quantify this aspect, we consider two distributions $\pi_{AA}(\epsilon; x, z)$ and $\pi_{VA}(\epsilon; x, z)$. $\pi_{AA}(\epsilon)$ is the probability of an $A$ site having $\epsilon$ $A$ first nearest neighbors. Similarly, $\pi_{VA}(\epsilon)$ is the probability of a vacant site $V$ with $\epsilon$ $A$ neighbors. $\epsilon$ refers to the local environment around an $A/V$ site. Here $\epsilon \in [0, c]$ and $c$ is the coordination number of the lattice. For a square lattice, $c$ has a value of 4.

Figure 1(c) and (d) show $\pi_{AA}$ corresponding to panels a and b, respectively. For $x = 0.2, z = 0.5$, $\pi_{AA}$ becomes stationary after 1 million additional iterations after reaching the target $z$. The configuration has attained complete equilibrium at this point. For $x = 0.2, z = 0.9$ it takes longer. No appreciable change in $\pi_{AA}$ is observed after an additional 2 billion iterations from the target structure. The lattice in Figure 1 (a-b) contains 105625 sites. Therefore, a disparity in the relaxation behavior is observed within the same system upon varying $z$. Since $\pi_{AA}$ and $\pi_{VA}$ are continuous functions of $x$ and $z$, this disparity can affect the calculation of thermodynamic properties[18–20] and care is needed to properly converge the probability distributions.

A movie showing the RMC evolution is provided as Supplementary Material. See Section 6 for details of the calculation. The starting arrangement is completely random. The movie shows the evolution starting from the point just before target has been achieved. In general, larger clusters are expected when $z$ is high – recall that $z$ is the probability of finding $A$ around an $A$-site. Clustering proceeds via two pathways, namely, the evaporation-condensation mechanism, which is predominant, and to a lesser extent by a diffusion-driven coarsening mechanism. The evaporation-condensation mechanism is witnessed from an early stage. Small clusters of $A$ begin to rapidly disappear and are added to existing large clusters. This aspect is visible in the movie. The clusters do not appear to execute a large-scale random walk, however, neck-formation between two clusters followed by coarsening can be seen occasionally. The lack of diffusion might be an outcome of only swap moves being attempted in our RMC calculations. Since the target $z$ is achieved, there is less propensity for rapid coarsening. Nonetheless, small clusters of $A$ do appear from time-to-time resulting in the system attaining some sort of a fast, dynamic equilibration. Figure 1(b) shows examples of such small clusters once target is reached (see target + 50 million). Similarly, small clusters of vacant sites can be seen within larger clusters of $A$. These clusters aid in the subsequent evolution and the evaporation-condensation mechanism continues to prevail. As the clusters become larger in size, it becomes difficult for them to fully undergo evaporation. The latter part of the evolution is slow.

## 2.3   Role of $\theta$ in RMC calculations

In Step 3 of the RMC algorithm (Section 2.1), $\theta$ dictates the average deviation in $N_{AA}$ in the converged structure with respect to $N_{AA,t}$. The effect of $\theta$ on $d_t^2$ is shown in Figure 2 for square lattice with $x = 0.2$ and $z = 0.5$. For all values of $\theta$ considered, $d_t^2$ rapidly decreases for the first ~0.1 million iterations. Thereafter, it stabilizes at a value and fluctuates around it. Moves that take the system closer to/away from the target are both accepted at this point. The value of $d_t^2$ increases with increasing $\theta$. Thus, $\theta$ is analogous to the temperature in a Metropolis MC calculation wherein moves resulting in large positive energy change are more likely to be accepted at higher temperatures.

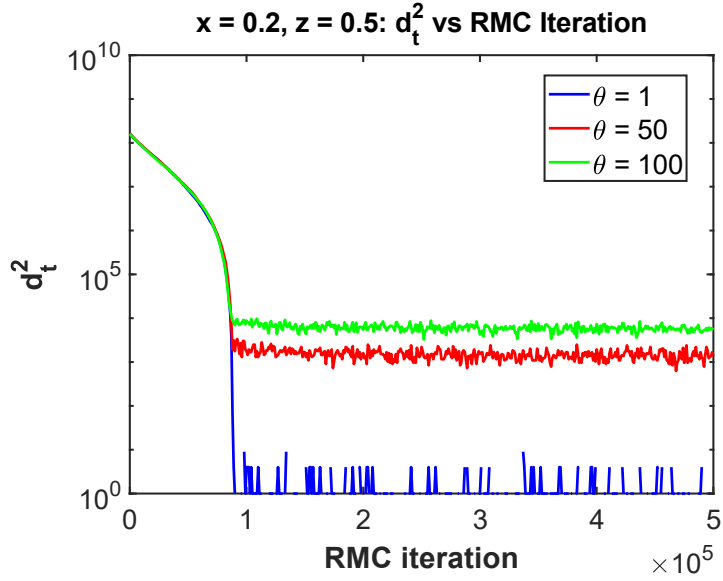

Figure 2: $d_t^2$ vs number of RMC trial moves for $x = 0.2$ and $z = 0.5$ in a square lattice. Results for $\theta = 1, 50, 100$ are shown.

The swap move in Step 2 of the RMC algorithm can be written as

$$A(\epsilon) + V(\epsilon') \rightleftharpoons A(\epsilon') + V(\epsilon). \tag{2}$$

Here $\epsilon$ and $\epsilon'$ are the number of 1NN $A$ neighbors for the selected $A$ and $V$ site in the move. Equation (1) can be rewritten as:

$$p_{acc} = \min\left[1, \exp\left(-\frac{(\epsilon'-\epsilon)^2 + 2(\epsilon'-\epsilon)(N_{AA}-N_{AA,t})}{\theta}\right)\right]. \tag{3}$$

$p_{acc}$ depends on $\theta$, the environments involved and the current value for $N_{AA}$. Suppose $N_{AA} > N_{AA,t}$, a trial move involving $\epsilon' > \epsilon$ will result in $N_{AA,n} > N_{AA}$, i.e., the RMC configuration moves away from the target. Such a move is associated with a small value of $p_{acc}$. For large values of $\theta$, $p_{acc}$ for such a trial move increases. This results in $d_t^2$ being higher. Similar arguments can be made when $N_{AA} < N_{AA,t}$. In fact, one can note a symmetry in $p_{acc}$ as shown in Figure 3. This symmetry arises based on the combination of $N_{AA} - N_{AA,t}$ and $\epsilon' - \epsilon$. Usually the probability of selecting a pair of sites such that $\epsilon' < \epsilon$ is different from the one for selecting $\epsilon' > \epsilon$. This causes the average $N_{AA}$ to systematically deviate from $N_{AA,t}$ in one direction – either above or below $N_{AA,t}$.

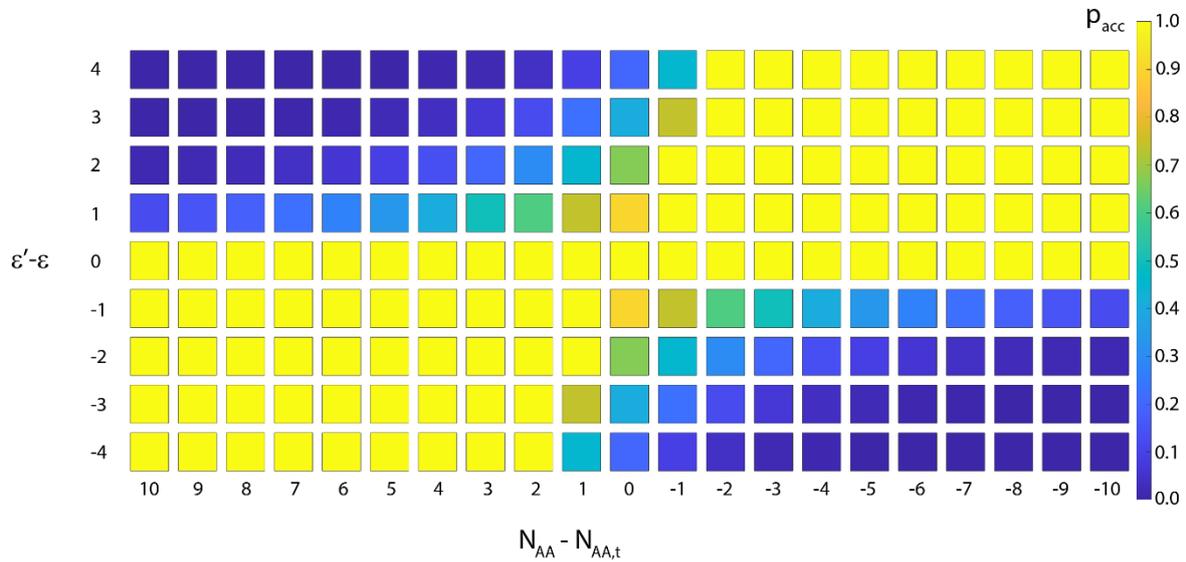

Figure 3. Colormap showing the acceptance probability $p_{acc}$ calculated using Equation (3) for $\theta = 10$.

From Equation (3) we infer that the average deviation for $N_{AA} - N_{AA,t} \propto \theta$. This trend is observed in Figure 2. For $\theta = 1, 50$ and $100$, the average value of $N_{AA} - N_{AA,t}$ is -0.74, -37.76 and -75.60, respectively. Historically, values of $\theta$ such as $\theta = 1$[13,16] and $\theta = 2$[26,27] have been used. We have used $\theta = 1$ in our present RMC studies.

## 3. RMC equilibrium criterion based on detailed balance

Equilibrium is said to be achieved when the forward and backward flux for the move shown in Equation (2) are equal. Equivalently, when a large number of trial moves are attempted at equilibrium, the average number of accepted moves in the forward direction equals the one in the backward direction. This provides the equilibrium condition for Equation (2). From now on, we write $\epsilon' = \epsilon + \delta$. The RMC system size is assumed to be large. Therefore, typically the selected $A - V$ pair are not neighbors. The probability of selecting the pair is given by $p(A(\epsilon), V(\epsilon + \delta)) = \pi_{AA}(\epsilon)\pi_{VA}(\epsilon + \delta)$. The detailed balance condition is:

$$\frac{\pi_{AA}(\epsilon)\pi_{VA}(\epsilon + \delta)}{\pi_{AA}(\epsilon + \delta)\pi_{VA}(\epsilon)} = \langle \frac{p_{acc}^{b\delta}}{p_{acc}^{f\delta}} \rangle. \quad (4)$$

The angular brackets in Equation (4) denotes averaging over trial moves. Alternatively,

$$\frac{\pi_{AA}(\epsilon)\pi_{VA}(\epsilon + \delta)}{\pi_{AA}(\epsilon + \delta)\pi_{VA}(\epsilon)} = K_\delta. \quad (5)$$

where

$$K_\delta = \langle \frac{\min\left(1, \exp\left(\frac{\delta^2 + 2\delta(N_{AA} - N_{AA,t})}{\theta}\right)\right)}{\min\left(1, \exp\left(-\frac{\delta^2 + 2\delta(N_{AA} - N_{AA,t})}{\theta}\right)\right)} \rangle. \quad (6)$$

The LHS of Equation (5) is the probability ratio for a given $\delta$, which is denoted as $PR_\delta$. The RHS of Equation (5) does not contain $\epsilon$. Thus, we arrive at any important conclusion: $PR_\delta$ calculated for any $\epsilon$ at equilibrium should be a constant $K_\delta$ that depends only on $\delta$. This point is not obvious by looking at the values of $\pi_{AA}(\epsilon)$ or $\pi_{VA}(\epsilon)$, for example see Figure 1c-d. The detailed balance equation helps uncover this relationship.

$K_\delta$ is an equilibrium constant for RMC. Equation (5) forms the basis for a metric that can be used to assess detailed balance condition in RMC. Since $\epsilon, \epsilon' \in [0, c]$ where $c$ is the 1NN coordination number, only certain values of $\delta$ are allowed. Moreover, $\pi_{AA}(\epsilon)$ and $\pi_{VA}(\epsilon)$ are independent of $\theta$. We infer that $K_\delta$ is independent of $\theta$.

## 4. Results and Discussions

### 4.1 Effect of $\theta$ on detailed balance and probability distributions

First, we demonstrate that Equation (4) is indeed satisfied at equilibrium. The LHS and RHS of Equation (4) is plotted against $\theta$ in Figure 3 with $x = 0.2$ and $z = 0.5$. Results are shown for $\theta = 1, 5, 10, 20, 50$ and $100$. Symbols show the $PR_\delta$ values, whereas lines represent $\langle \frac{p_{acc}^{b\delta}}{p_{acc}^{f\delta}} \rangle$. Only the converged $\pi_{AA}$ and $\pi_{VA}$ are used to calculate $PR_\delta$. After convergence, an additional 0.1 million trial moves are used to calculate $p_{acc}^{f\delta}$ and $p_{acc}^{b\delta}$. As shown, $PR_\delta$ does not vary with $\theta$. In addition, we find that RHS of Equation (4) matches with $PR_\delta$ for $\delta = 1, 2$ irrespective of $\theta$. For large $\delta$, the RHS shows slight deviation from $PR_\delta$. This can be explained in terms of the sampling error in $\langle \frac{p_{acc}^{b\delta}}{p_{acc}^{f\delta}} \rangle$ – since the probability of selecting the pair $A(\epsilon) - V(\epsilon + \delta)$ is small for large $\delta$, fewer moves are attempted. Overall, Figure 3 confirms detailed balance in a converged RMC structure for all $\delta$ and $\theta$.

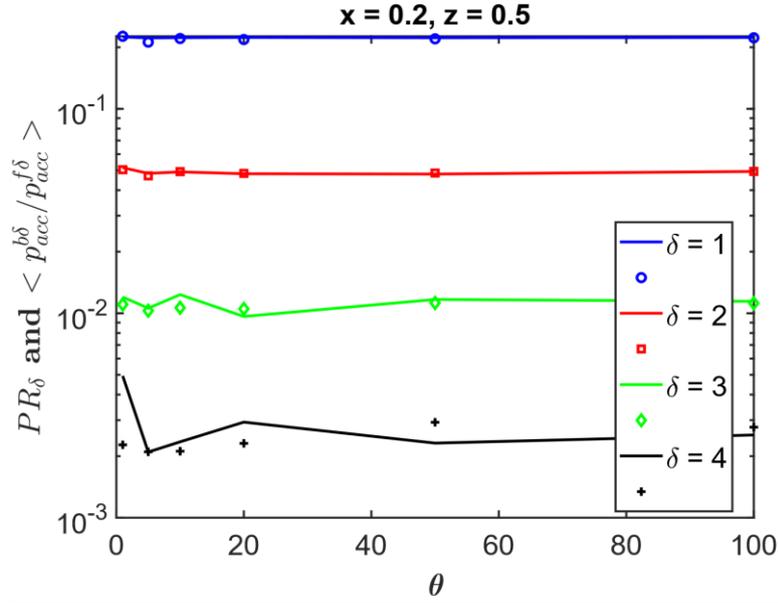

Figure 3: Variation of LHS and RHS of the detailed balance equation (Equation (4)) with respect to $\theta$ for different values of $\delta$. Lines represent the $\langle \frac{p_{acc}^{b\delta}}{p_{acc}^{f\delta}} \rangle$, whereas symbols represent $PR_\delta$. Results are shown for $x = 0.2$ and $z = 0.5$.

Using Equation (5),

$$K_2 = K_1^2, \qquad (7)$$

$$K_3 = K_1^3,$$

and so on. This shows that $K_1$ is the only independent equilibrium constant. Other equilibrium constants can be derived from $K_1$.

From Figure 2 it is evident that $d_t^2$ decreases almost identically for different $\theta$ as the RMC calculation approaches the target constraints. The subsequent relaxation is shown in Figure 4 in terms of $\pi_{AA}(\epsilon)$ versus iterations for $x = 0.2$ and $z = 0.9$. Symbols and lines denote the results obtained with $\theta = 1$ and 100, respectively. Results are shown for 10, 20 and 100 million iterations. Interestingly, after reaching target the relaxation dynamics still does not depend on $\theta$. This suggests that there is no particular advantage in selecting a particular value of $\theta$.

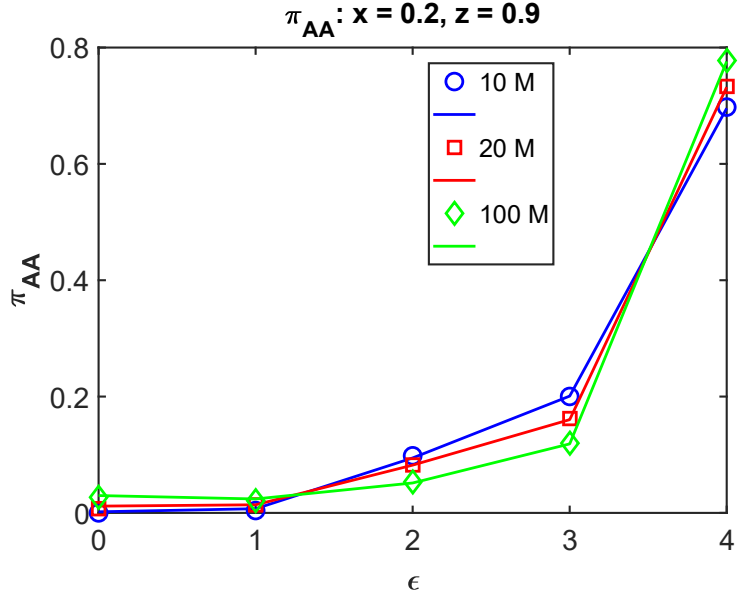

Figure 4: Variation in $\pi_{AA}(\epsilon)$ with iterations for $x = 0.2$, $z = 0.9$ in case of square lattice. Symbols and lines represent probability distribution obtained using $\theta = 1$ and $100$.

### 4.2 Metric for *on-the-fly* assessment of the detailed balance condition

From our previous discussion, we find that for any value of $\delta$:

$$\frac{\pi_{AA}(\epsilon)\pi_{VA}(\epsilon + \delta)}{\pi_{AA}(\epsilon + \delta)\pi_{VA}(\epsilon)} = K_1^\delta. \tag{8}$$

Here $\delta$ is present as the exponent in the right-hand side of the equation. Figure 5 shows the evolution of $PR_\delta$ for $x = 0.2, z = 0.5$. The vertical dashed line in each panel indicates the point where the target structure is first achieved. At the beginning of calculation $PR_\delta$ is 1 for all $\delta$. This is because the initial arrangement is perfectly random. In such a case, $\pi_{AA}(\epsilon)$ and $\pi_{VA}(\epsilon)$ are identical; $\pi_{AA}(\epsilon)$ and $\pi_{VA}(\epsilon)$ are a binomial distribution. In Figure 5a, the probability ratio is shown for different combinations of $\epsilon$ and $\epsilon'$ while $\delta = 1$. Deviation in $PR_{\delta=1}$ values can be quite significant when the target is reached and remains visible till ~0.6 million. The probability used in the calculation of $PR_{\delta=1}$ is averaged using 128 independent RMC calculations to lower the statistical noise in small probability values. As expected, eventually all $PR_\delta$ values are equal to $K_1^\delta$ indicating detailed balance is satisfied. At this point, the probability distributions do not show any systematic change. For sake of completeness, Figure 5b-d shows the behavior for $PR_\delta$, $\delta = 2 - 4$.

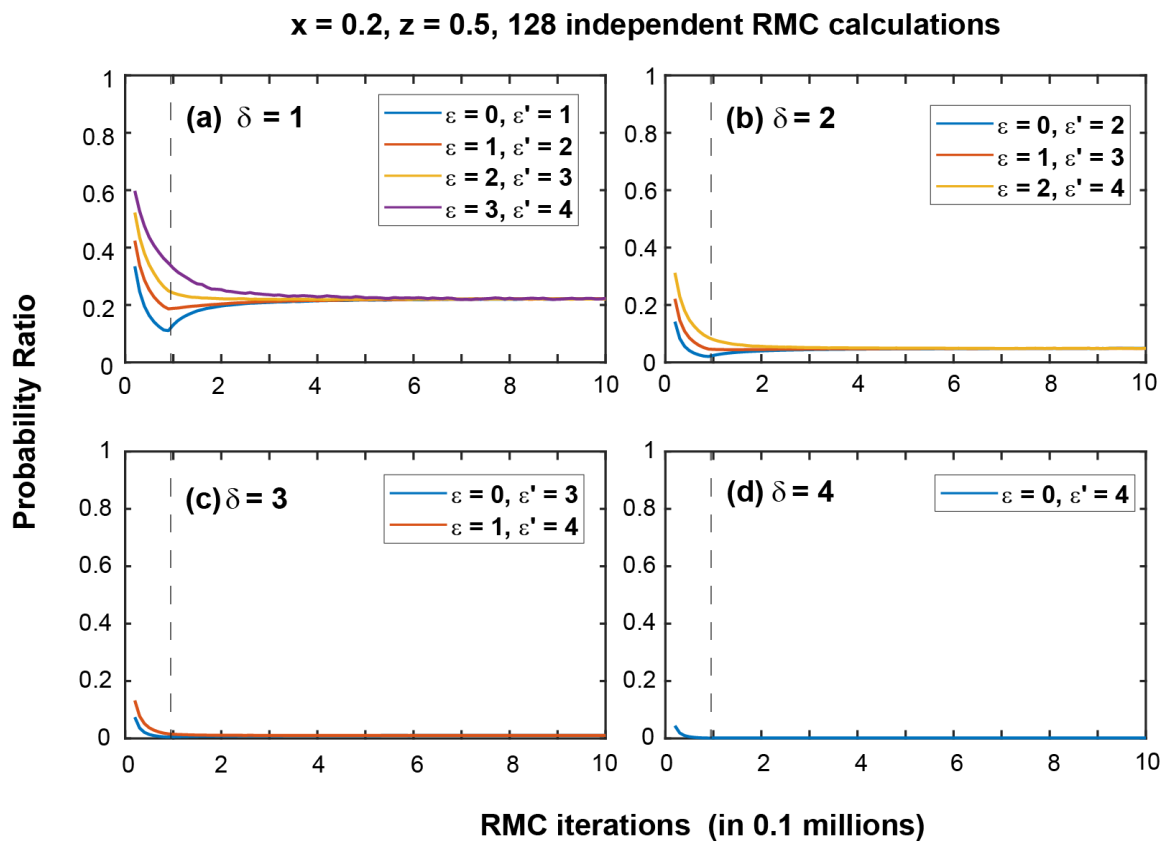

Figure 5: $PR_\delta$ vs RMC iteration for $x = 0.2$ and $z = 0.5$. Dashed vertical line represents the point at which the target structure is achieved.

Figure 6 shows the evolution of $PR_\delta$ for a different condition, namely, $x = 0.2$ and $z = 0.9$, where we have earlier seen that the relaxation can be very slow. Several aspects are similar to the ones in Figure 5. See Figure 6a for example. Initially, a large variation in the values for $PR_{\delta=1}$ is observed. The variation is present at the time the target is first reached. Finally, the $PR_\delta$ curves overlap indicating detailed balance is satisfied. However, $PR_\delta$ continues to change with trial moves because of the slow relaxation in the quasi-equilibrated RMC structure. The horizontal-axis is shown in log-scale to highlight this observation.

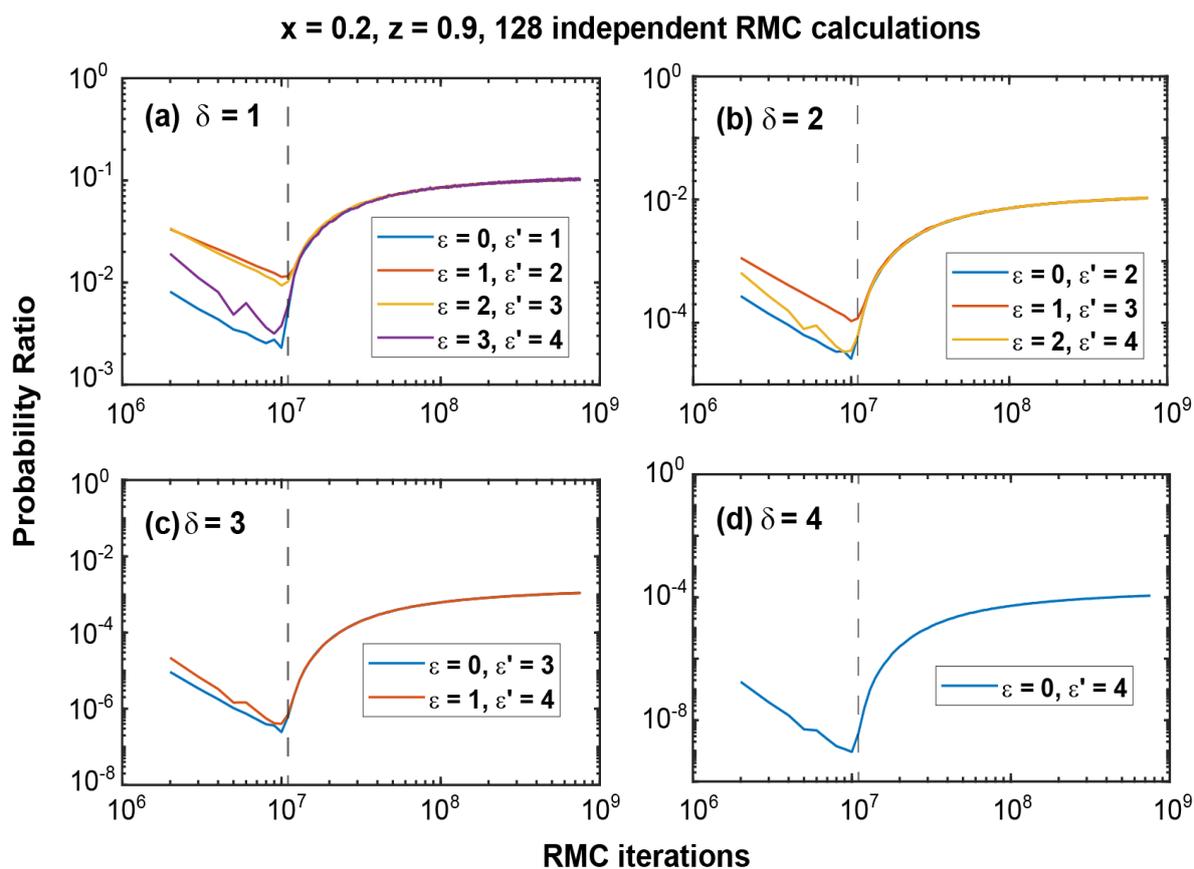

Figure 6: $PR_\delta$ vs RMC iteration for $x = 0.2$ and $z = 0.9$. Dashed vertical line represents the point at which the target structure is achieved.

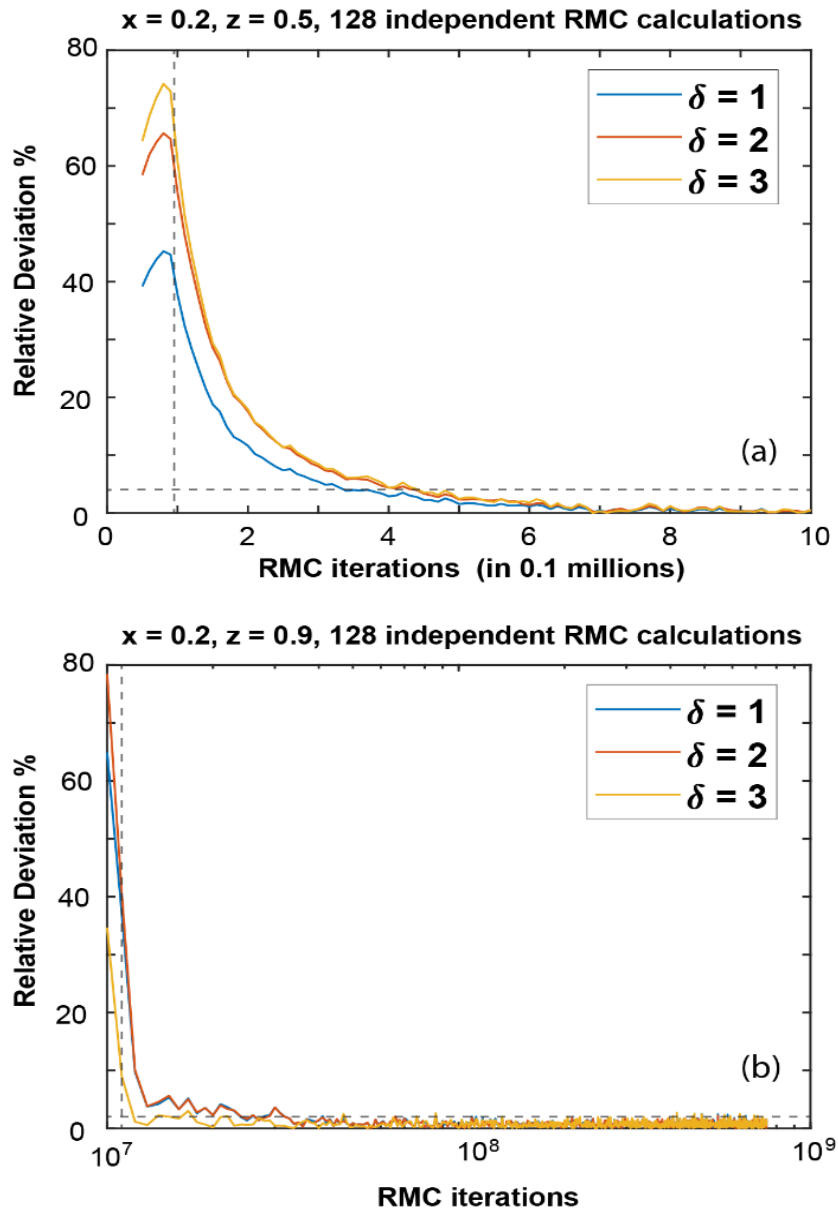

Figure 7: Relative deviation (in %) vs RMC iterations for (a) $x = 0.2, z = 0.5$ and (b) $x = 0.2, z = 0.9$ in case of a square lattice.

To assess whether quasi-equilibrium is achieved with the current RMC configuration we introduce a metric that involves calculation of the relative deviation in $PR_\delta$ (when different $\epsilon$ values are possible):

$$Relative\ deviation = \frac{Standard\ deviation\ of\ PR_\delta}{Mean\ of\ PR_\delta} \times 100\ \%. \tag{9}$$

Figure 7a shows the relative deviation as defined in Equation (9) with respect to RMC trial moves for $x = 0.2$ and $z = 0.5$. For all values of $\delta$, the relative deviation decreases rapidly till 0.8 million iterations and thereafter it fluctuates around a mean value. After 0.6 million iterations the relative deviation for $\delta = 1, 2$ and 3 are found to be 3.27 %, 3.32 %, and 0.8 %, respectively. At this point, the RMC calculations are deemed to have satisfied the detailed balance. In general, we suggest that the RMC calculations can be stopped and the probability values can be taken as converged when the relative deviation is less than a user-defined tolerance value (for example, 5 %) and slow relaxation is absent. For $x = 0.2$ and $z = 0.9$, ~10 million iterations are required to achieve quasi-equilibrium. Relative deviation for $\delta = 1, 2$ and 3 is found to be 2.28 %, 3.11 %, and 1.95 % respectively. An approach to measure slow relaxation is discussed next.

### 4.3 Extension to problems involving multiple SRO parameters

Often RMC applications involve the use of multiple SRO parameters for construction of 2D and 3D structures. For instance, the rdf specifies the pair probability in multiple neighbor shells. This is equivalent to specifying the SRO parameters for the first nearest neighbors, second, and so on. Here we study the relaxation behavior when two SRO parameters are involved.

Consider the SRO parameters $z_1$ and $z_2$, which denote the probability of finding $A$ in the 1st and 2nd coordination shell, respectively, around an $A$-site. Analogous to Equation (2), the swap move can be written as

$$A(\epsilon_1, \epsilon_2) + V(\epsilon_1', \epsilon_2') \rightleftharpoons A(\epsilon_1', \epsilon_2') + V(\epsilon_1, \epsilon_2). \tag{10}$$

$\epsilon_1$ and $\epsilon_2$ are the number of $A$ neighbors around the selected $A$-site, while $\epsilon_1'$ and $\epsilon_2'$ are the number of $A$ neighbors around the selected $V$-site for the move in the forward direction. Subscripts 1 and 2 indicate the first and second neighbor shell. The probability of finding such an $A$ site is $\pi_{AA}(\epsilon_1, \epsilon_2)$ whereas the probability of finding such a $V$ site is $\pi_{VA}(\epsilon_1', \epsilon_2')$. For square lattice $\epsilon_1, \epsilon_2 \in [0,4]$. We define the distance from the target as

$$d_t^2 = (N_{AA,1} - N_{AA,t,1})^2 + (N_{AA,2} - N_{AA,t,2})^2. \tag{11}$$

The detailed balance condition is

$$\frac{\pi_{AA}(\epsilon_1, \epsilon_2)\pi_{VA}(\epsilon_1', \epsilon_2')}{\pi_{AA}(\epsilon_1', \epsilon_2')\pi_{VA}(\epsilon_1, \epsilon_2)} = \langle \frac{p_{acc}^{b\delta_1\delta_2}}{p_{acc}^{f\delta_1\delta_2}} \rangle. \tag{12}$$

The right-hand side is a function of $\delta_1 = \epsilon_1' - \epsilon_1$ and $\delta_2 = \epsilon_2' - \epsilon_2$. Therefore, the same value of the probability ratio can be attained with different $\epsilon_1, \epsilon_2$ but fixed $\delta_1, \delta_2$. It follows that

$$PR_{0,\delta_2} \, PR_{\delta_1,0} = PR_{\delta_1,\delta_2}. \tag{13}$$

For this reason, it suffices to focus specifically on the convergence of $PR_{0,1}$ and $PR_{1,0}$ since all other probability ratios are obtained from these two. The corresponding equilibrium constants are $K_{0,1}$ and $K_{1,0}$.

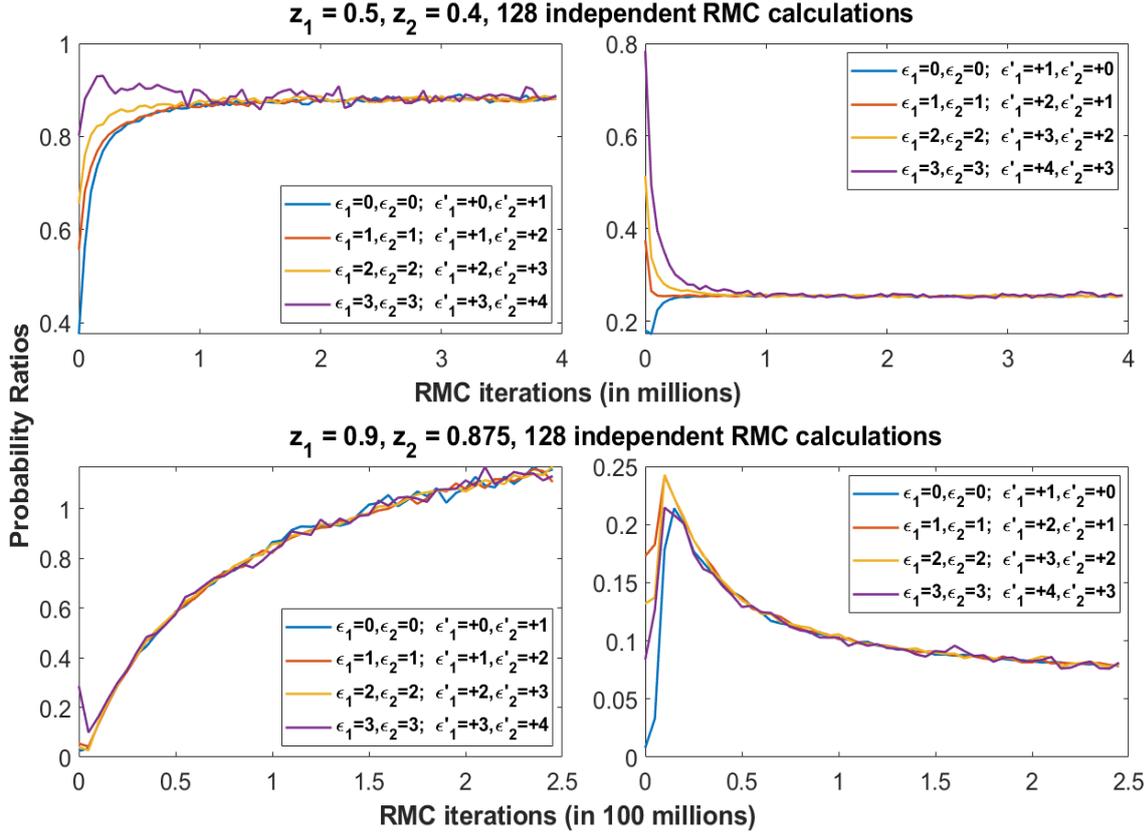

Figure 8: Top row: Probability ratio for $x = 0.2, z_1 = 0.5, z_2 = 0.4$. Bottom row: $x = 0.2, z_1 = 0.9, z_2 = 0.875$. For left column, $\delta_1 = 0, \delta_2 = 1$, whereas for right column, $\delta_1 = 1, \delta_2 = 0$.

Figure 8 shows the probability ratio for $x = 0.2$ with different combinations of $\epsilon_1$ and $\epsilon_2$ for given values of $\delta_1$ and $\delta_2$. The behavior is similar to a single SRO parameter. Convergence is achieved more easily with $z_1 = 0.5, z_2 = 0.4$ compared to $z_1 = 0.9, z_2 = 0.875$. The former case converges in 2.5 million RMC iterations while the latter converges only after 500 million trial moves. Latter RMC calculations are computationally expensive and require several hours for completion.

The relaxation behavior in RMC appears analogous to the one observed in glasses and polymers[28–33]. The dynamics of supercooled liquids such as glass is characterized by two or more relaxation timescales. In Figure 8, the probability ratios appear to converge to a common value, as required by detailed balance, and then proceed to continue to relax further while the structure is quasi-equilibrated. This observation that the detailed balance condition is satisfied can be termed as fast relaxation. Further evolution of the probability ratios resembles slow relaxation in glasses. The convergence happens monotonically as seen in Figure 8. The relaxation dynamics of glassy structures with time has been described by the Kohlrausch-William-Watts (KWW) function, also known as the stretched exponential function[31]. We employ a similar functional form here. We assume that for large values of $z$ the probability value $f$ can be described in terms of number of trial moves as

$$f = f_0 \left[1 + c\left(1 - \exp\left(-\left(\frac{\Delta I}{I_0}\right)^\alpha\right)\right)\right]. \tag{14}$$

Here $f$ can be $\pi_{AA}(\epsilon_1, \epsilon_2)$ or $\pi_{VA}(\epsilon_1, \epsilon_2)$, $\Delta I$ is the number of trial moves after reaching target and quasi-equilibrium, $\Delta I \geq 0$, $I_0$ is a characteristic number of RMC trial moves. $f_0$ is the probability value when $\Delta I = 0$ and $\alpha$ is an exponent. Typically, for stretched exponent $\alpha < 1$. The probability finally converges to

$$\hat{f} = f_0(1 + c) \tag{15}$$

$c$ is a measure of the distance from complete convergence once the system is quasi-equilibrated for the first time.

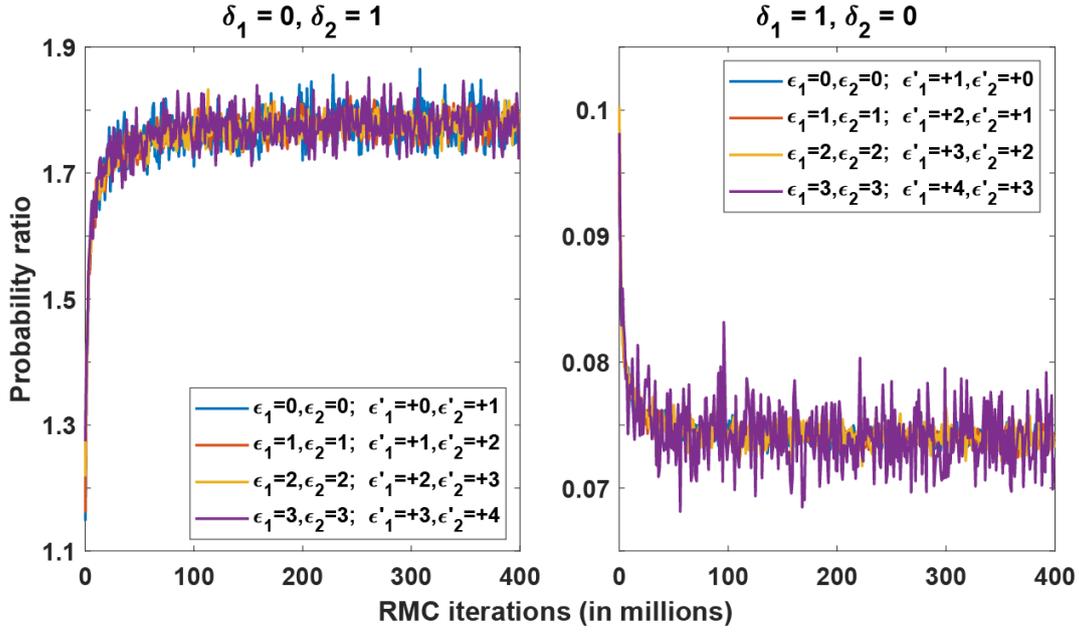

Figure 9: Probability ratio $\frac{\pi_{AA}(\epsilon_1,\epsilon_2)\pi_{VA}(\epsilon'_1,\epsilon'_2)}{\pi_{AA}(\epsilon'_1,\epsilon'_2)\pi_{VA}(\epsilon_1,\epsilon_2)}$ versus RMC trial moves for $x = 0.2$ $z_1 = 0.7$ and $z_2 = 0.6$. Here $\delta_1 = \epsilon'_1 - \epsilon_1$ and $\delta_2 = \epsilon'_2 - \epsilon_2$. Dashed vertical line indicates when the target structure is first reached. A total of 100 million RMC trial moves are attempted.

Consider the RMC simulation of a configuration for $x = 0.2, z_1 = 0.7$ and $z_2 = 0.6$ as shown in Figure 9. In this case, the system appears to be quasi-equilibrated well before the target is reached. Therefore, $f_0 = f_{target}$ where $f_{target}$ denotes the probability value when target is first reached. For the slow relaxation part, we fit the stretched exponential function (Equation (14)) to the data collected. This example is chosen because the calculation is properly converged in a few 100 million iterations. From Figure 10, it is evident that the KWW function can be used to characterize the slow relaxation. Each panel shows the measured probability value from a single RMC calculation in symbols along with the corresponding fitted KWW function in line. Each of the panels have $R^2 > 0.95$, indicating very good agreement.

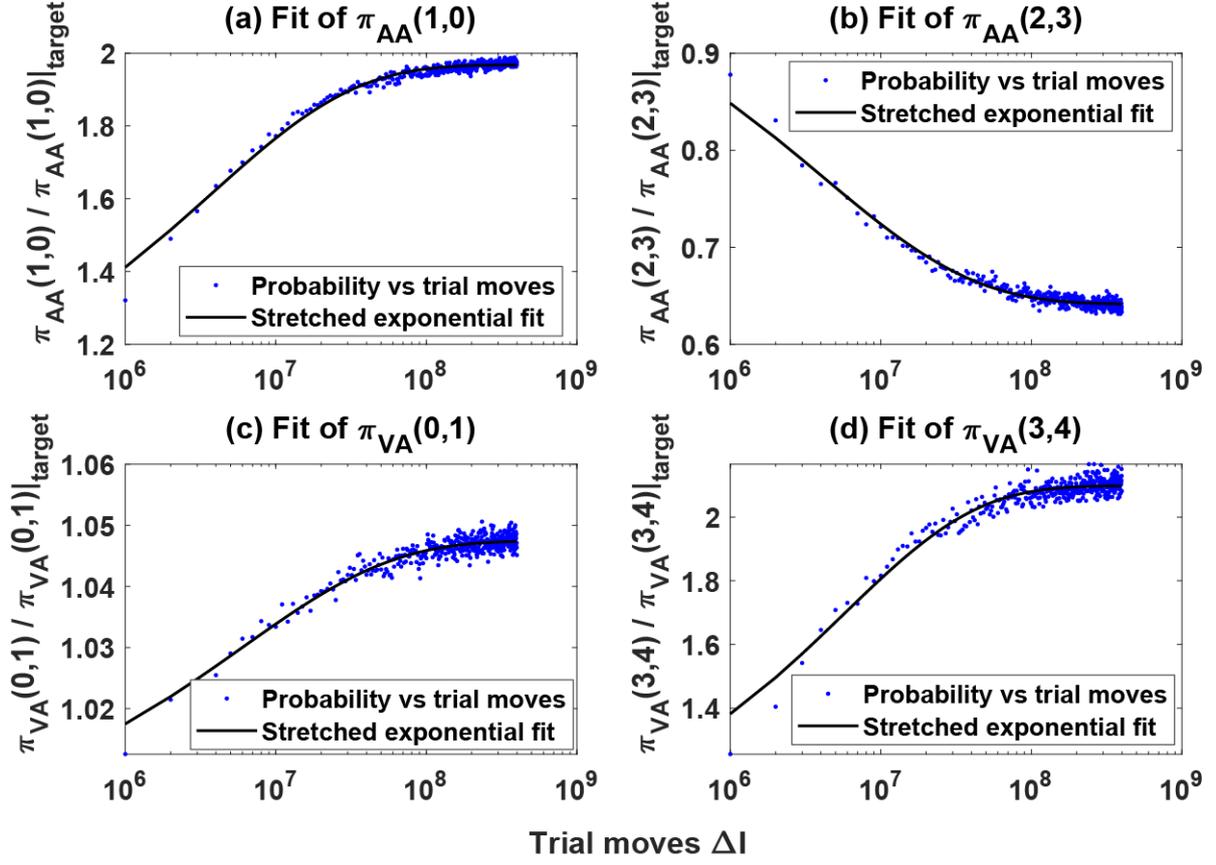

Figure 10: $\pi$ from RMC calculations (in symbols). Line shows the fitted stretched exponential function. Fitted parameters are: (a) $I_0 = 3.692 \times 10^6, c = 0.968, \alpha = 0.4523$, (b) $I_0 = 4.063 \times 10^6, c = -0.2586, \alpha = 0.4274$, (c) $I_0 = 5.98 \times 10^6, c = 0.04747, \alpha = 0.4331$, (d) $I_0 = 5.64 \times 10^6, c = 1.098, \alpha = 0.4912$.

Figure 10 caption shows the fitted parameter values. It is interesting to note that although the RMC data appears to converge after few 100 million trial moves, yet $I_0$ is two orders of magnitude smaller and lies between 3-6 million RMC moves. The value of $\hat{f}$ is obtained from Figure 10 with the help of Equation (15). The number of RMC moves required so that in Equation (14) $\frac{f-f_0}{\hat{f}-f_0} = 0.99$ is

$$\Delta I = (4.606)^{1/\alpha} I_0. \tag{16}$$

From this analysis, we estimate that it takes 100-300 million iterations to converge.

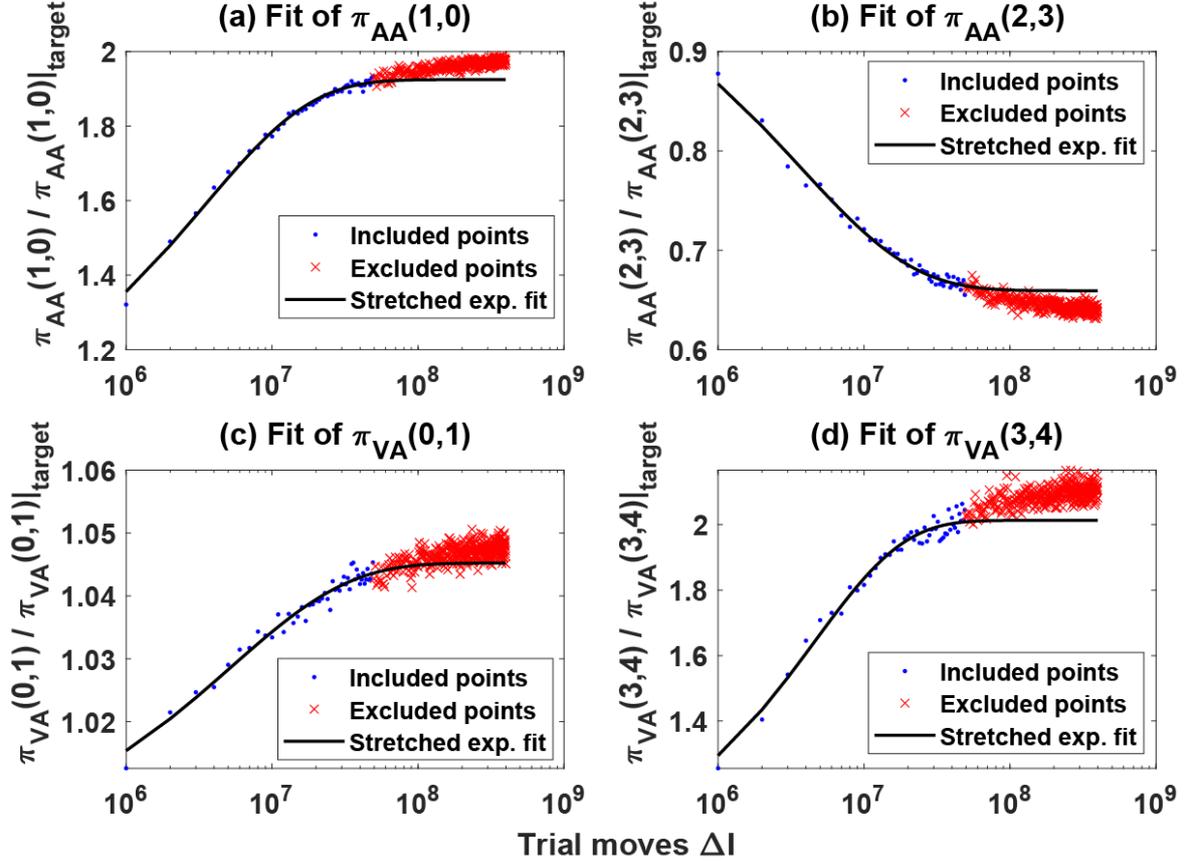

Figure 11: RMC probabilities used for fitting the stretched exponential function are shown in blue dots. Red cross denotes $\pi_{AA}$ data from $\Delta I$=50 to 400 million trial moves. Stretched exponential (black line) is fitted to $\pi_{AA}$ data till $\Delta I$= 50 million trial moves. Fitted parameter values are: (a) $I_0 = 3.401 \times 10^6, c = 0.9248, \alpha = 0.5904$, (b) $I_0 = 3.611 \times 10^6, c = -0.3407, \alpha = 0.5522$, (c) $I_0 = 5.185 \times 10^6, c = 0.04524, \alpha = 0.5336$, (d) $I_0 = 4.546 \times 10^6, c = 1.013, \alpha = 0.7054$.

Next, we explore the possibility of predicting $\hat{f}$ without attempting excessively long RMC calculations. This feature can be particularly useful for large $z_{AA}$. Such an exercise is performed in Figure 11 with the data taken from Figure 10. We require target constraints and quasi-equilibrium to be achieved. Some additional RMC moves are considered so that a numerical fit to the probability data is obtained. Panels in Figure 11 show the KWW function fitted using data from $\Delta I$=0 to 50 million trial moves for different $\pi$ values. Finally, one uses the KWW function to predict the converged probability value. The fitted parameters are shown in caption of Figure 11. Applying Equation (16) to these parameter values, we conclude that it takes about 100 million iterations to converge. Therefore, the relaxation in the original RMC data is slower than the one obtained using the fitted KWW function in Figure 11. When the fitted KWW function is extrapolated to $\Delta I$=400 million trial moves, it is observed that the extrapolated values follow the original RMC data (red cross symbol), however, there is a gap between the original data and extrapolated results. The error in the converged probability value $\hat{f}'$ obtained from extrapolation is calculated.

Table 1. Error in RMC predictions when the calculation is stopped immediately after fast relaxation has been achieved. Parameters are identical to the ones in Figure 10.

| Panel | $\hat{f}$ | Standard practice | | Extrapolated probability | |
|---|---|---|---|---|---|
| | | $f_0$ | Error (%) in $f_0$ | $\hat{f}'$ | Error (%) in $\hat{f}'$ |
| a | 0.059976 | 0.030468 | 49.2 | 0.05842 | 2.59 |
| b | 0.018759 | 0.029618 | 57.9 | 0.019676 | 4.89 |
| c | 0.107636 | 0.102988 | 4.3 | 0.107478 | 0.15 |
| d | 0.001929 | 0.000927 | 51.9 | 0.001861 | 3.53 |

Two types of errors are shown in Table 1. First, suppose $f_0$ was reported as the probability value from RMC, the error $\left|1-\frac{f_0}{\hat{f}}\right|$ is proportional to $|c|$, i.e., the height or depth of the plateau region in Figure 10. This error arises from stopping the RMC calculation immediately after fast relaxation has been achieved – this corresponds to standard practice. Table 1 demonstrates that the error can be as large as 50%. Alternatively, if we consider fitting the KWW function to limited RMC data as in Figure 11, the error is $\left|1 - \frac{\hat{f}'}{\hat{f}}\right|$. Table 1 shows that the error in this case is much lower than the error in $f_0$. The largest error is found to be less than 5%. Thus, the KWW function helps in (i) describe the convergence of RMC, (ii) estimating the number of trial moves required to converge the calculation, and (iii) reduce the number of RMC iterations required for estimating the probability distributions especially with large $z_{AA}$ by extrapolating the fitted KWW function.

## 5. Conclusion

The reverse Monte Carlo (RMC) technique is used extensively in literature for creating atom-based structural models. The usual practice in RMC is to stop the calculation once target constraints are reached, however, this may lead to an incompletely converged structure. This study provides new insights into the dynamic relaxation in RMC and issues related to convergence behavior. In RMC, equilibration can consist of an initial fast relaxation followed by a slow relaxation. Fast relaxation can be probed by assessing how well detailed balance condition is satisfied. In such a case, the dynamics can be collapsed into probability ratios that are used for determining fast relaxation. Concepts such as equilibrium constants are developed to relate the different probability ratios. We find that fast relaxation is achieved soon after the target constraints are reached. Slow relaxation, on the other hand, happens on a comparatively longer timescale. The slow dynamics can be described using a stretched exponential function. This aspect can be exploited to calculate error in probability distributions from RMC and even to predict the converged values. We believe that these insights are useful to users of the RMC method and provides them with a mathematical framework to probe the relaxation dynamics of RMC. An emphasis on the convergence aspects of RMC is essential for the reliability and reproducibility of results.

## 6. Supplementary Material

The movie available as a Powerpoint presentation shows relaxation dynamics in reverse Monte Carlo (RMC) calculations. Two SRO parameters are used. RMC parameters are: $x = 0.2, z_1 = 0.9, z_2 = 0.85$ and $\theta = 1$. Total number of trial moves are 200 million. First frame is obtained at 0.5 million iterations. The target structure is reached at 7 seconds into the movie.

## 7. Data availability statement

Data will be provided upon request to authors.

## 8. Acknowledgements

AC acknowledges support from Science and Engineering Research Board, Grant Nos. EMR/2017/001520 and MTR/2019/000909, and National Supercomputing Mission DST/NSM/R&D_HPC_Applications/2021/02.